\documentclass[12pt]{article}
\usepackage{latexsym}

\newcommand{\lbl}[1]{\label{eq:#1}}

\newcommand{\be}{\begin{equation}}
\newcommand{\ee}{\end{equation}}
\newcommand{\bea}{\begin{eqnarray}}
\newcommand{\eea}{\end{eqnarray}}

\newcommand{\noi}{\noindent}
\newcommand{\nn}{\nonumber}

\newcommand{\lesssim}{ {\
\lower-1.2pt\vbox{\hbox{\rlap{$<$}\lower5pt\vbox{\hbox{$\sim$}}}}\ } 
}
\newcommand{\gtrsim}{ {\
\lower-1.2pt\vbox{\hbox{\rlap{$>$}\lower5pt\vbox{\hbox{$\sim$}}}}\ } 
}

\newcommand{\cO}{{\cal O}}

\newcommand{\QCD}{$QCD_{\infty}$}

\usepackage{epsfig}

\begin{document}

\begin{titlepage}

\begin{flushright} UAB-FT-502\\ 
\end{flushright}
\begin{center} 
{\Large \bf Large-$N_c$ QCD meets Regge theory:\\[0.5cm]
the example of spin-one two-point functions }\\[3.0cm]

{\bf Maarten Golterman}$^a$ and {\bf Santiago Peris}$^b$\\[1cm]

$^a$Department of Physics, Washington University, 
St. Louis, MO 63130, USA\\
e-mail: maarten@aapje.wustl.edu\\[0.5cm]
$^b$Grup de F{\'\i}sica Te{\`o}rica and IFAE\\ Universitat Aut{\`o}noma
de Barcelona, 08193 Barcelona, Spain\\
e-mail: peris@ifae.es

\end{center}

\vspace*{1.0cm}

\begin{abstract}

We discuss the phenomenological implications of assuming a Veneziano-type
spectrum for the vector and axial-vector two-point functions in QCD at large
$N_c$. We also compare the phenomenological results with those of 
Lowest-Meson Dominance, and find that they agree.    
\end{abstract}

\end{titlepage}

\section{Introduction}
\lbl{sec:int}
 
\noi

Quantum Chromodynamics simplifies when we take the number of colors, $N_c$,
to infinity \cite{thooft}; we will denote this limit by \QCD . 
Assuming confinement, it is generally believed that the large-$N_c$
solution qualitatively resembles the real world enough to be taken as a
serious starting point for a systematic approximation by expanding 
in inverse powers of $N_c$ \cite{witten}. 

Unfortunately, no solution to \QCD\ 
has been found to date, except in the two-dimensional case \cite{thoofttwo}. 
We know that even in the large-$N_c$ limit this
solution has to be very complex.  Large $N_c$, together with confinement
and asymptotic freedom, predicts the existence 
of an infinite set of zero-width meson resonances in any channel with
given quantum numbers.  These resonances  contribute to any Green's 
function made of quark bilinear operators corresponding to those quantum
numbers. At very large energies a simplification occurs, as the 
infinite tower of
resonances has to become equivalent to the free-quark loop of the parton
model obtained in leading-order perturbative QCD (pQCD), with the 
connection provided by the operator product expansion (OPE).
However, how this happens is not known, except in the two-dimensional case.
We would therefore like to propose a simple model for this equivalence,
which despite its simplicity is phenomenologically quite successful.

In this work we will deal only with two-point 
functions in the chiral limit. 
We limit ourselves to this choice because these are the simplest 
Green's functions which obey non-trivial chiral symmetry and OPE constraints. 
The model is based on the observation that, although the equivalence
between the resonance description of \QCD\ and the quark-gluon description of
pQCD is valid at asymptotically large energy scales,
phenomenology suggests that asymptotic behavior may set in already at values of
$s=Q^2\sim (1-2)\ {\rm GeV^2}$.  In the vector channel, the $\rho$ meson
is clearly lighter than this scale $s$, but all higher resonances
($\rho',\rho'',\rho''', $ {\it etc.}) appear to
lie already in this asymptotic region. 
Therefore, it is reasonable to consider this 
channel as consisting of a separate $\rho$ resonance of mass
$M_{\rho}$, plus an infinite tower labeled by an index $n$, of masses $M_V(n)$,
which corresponds to the perturbative description in a way to be described
below.
In the axial-vector channel, on the contrary, the $a_1$ may already be heavy
enough to group all resonances 
together in a corresponding tower, $M_A(n)$.  Here the exception is of
course the 
Goldstone meson which, because of chiral-symmetry breakdown, is massless 
and stays away from the tower. 

How these resonances are spaced as a function of the index $n$ is not known 
in \QCD~.  
Instead, we will rely on Regge theory \cite{collins}.
It has been long suspected that there is  a connection
between large-$N_c$ and Regge theory \cite{witten, thooft}, but the precise
link has never been found \cite{patel}.  
One of the natural features of Regge theory 
is the existence of so-called daughter trajectories. Their most
important feature for us is that they give rise
to towers of resonances with a given spin 
obeying an equal-spacing rule with respect to the index
$n$, {\it i.e.} the mass squared is given by $M^2(n) = a + b n$. 
In two dimensions, where the solution to \QCD\ is known, it can be shown that  
indeed $M^2(n)\sim n$ for large $n$. Here, in four dimensions, we will
conjecture that this linear behavior in $n$ is also valid  
for both the vector and axial-vector towers, and assume that 
\be
\label{spectrum}
M_{V,A}^2(n) = m_{V,A}^2 + n \ \Lambda_{V,A}^2 \, ,
\ee  
is a reasonable approximation 
for all mesons except the $\rho$ meson and the pion.

For simplicity, we will content ourselves with this equal-spacing (ES)
 ansatz producing only the leading term in pQCD,
the parton-model logarithm.  
A possible remedy to this is to delay the onset
of the equally spaced tower by including more and more individual resonances
which are not part of the tower, and thus move the ``onset" of this
tower to a higher energy, where the parton-model logarithm is a better
approximation.  In this way the description would become more 
accurate.  Alternatively, one could take the $n$ dependence in 
eq.~(\ref{spectrum}) to be more complicated,\footnote{In principle one could
 also play with the $n$ dependence of the resonances' decay constants (see
 below).} such that non-leading 
perturbative corrections are reproduced from our  ansatz as well.
The price to pay, of course, is that this introduces more unknown 
parameters into our ansatz, which would have to be fixed by
matching to the OPE and chiral perturbation theory (ChPT) to higher orders
in a way similar to what we will describe for our simple ansatz
in section 2. 

Much of this has been said before, although perhaps
not within the unifying framework of the large-$N_c$ expansion. For instance,
in the pioneering work by Bramon, Etim and Greco \cite{bramon} and Sakurai
\cite{sakurai}, it was already observed that a 
resonance tower {\it including} as the first state the $\rho$, and obeying 
\be
\label{GVMD}
M_V^2(n) = M_{\rho}^2 \left(1+ \alpha \ n \right)\ ,
\ee
reproduces reasonably well the parton-model logarithm of the two-point vector 
correlation
function if $\alpha = 2 $ and the $\rho$ decay constant is taken at its
experimental value. This was an exciting result because the Veneziano model
\cite{veneziano}, 
which can be considered as the model paradigm of Regge theory, 
predicted exactly this value.  We will later on see that, if the 
distribution of eq. (\ref{GVMD}) is assumed, the value $\alpha=2$ is actually 
an exact algebraic consequence of the OPE, independent of the value of 
the $\rho$ decay constant.   
However, in that case eq.~(\ref{GVMD}) also leads to
a negative value for the gluon condensate, a result which is clearly disfavored
phenomenologically \cite{yndurain}.  In our approach we will not take
the $\rho$ to be part of the equally spaced tower.

Also Geshkenbein  
considered \cite{geshkenbein} the problem of an infinite set of
resonances in the vector channel and how this would match onto the OPE; this
time with a general function $M^2_V(n)$. 
However, nobody seems to have considered an {\it infinite} set of resonances
in the vector and axial-vector
correlation functions on the same footing, together with their 
properties under chiral symmetry \cite{KdeR}.  In addition,
the experimental knowledge of
the vector spectrum has improved considerably recently, so that 
an updated comparison is interesting.  

In this paper, we will study the phenomenological consequences of
our ansatz, in which we take equally spaced towers in both the
vector and axial-vector channels with separate resonances for the $\rho$
and the pion, to leading order in $N_c$, and in the chiral limit. Furthermore,
although we will refer explicitly to the $\rho$ and $a_1$ mesons for
simplicity, it should be clear that our results also apply in the $U(3)\times
U(3)$ chiral limit. 

The rest of the paper is organized as follows. 
Section 2 is devoted to the presentation of our model. Section 3 contains the
phenomenological consequences of this model. In section 4 we make a comparison
with the Lowest-Meson Dominance approximation of ref. \cite{previouswork},
and we conclude in section 5.
   
\section{The model}

Let us start by defining the vector two-point function as 
\be
\label{correlator}
\Pi_{V,A}^{\mu\nu}=\ i\,\int d^4x\,e^{iqx}\langle J_{V,A}^\mu(x) 
J_{V,A}^{\dag \nu}(0)\rangle
= \left(q_{\mu} q_{\nu} - g_{\mu\nu} q^2 \right)\Pi_{V,A}(q^2) \ ,
\ee
with $J_{V}^\mu(x) = {\overline d}(x)\gamma^\mu u(x)$ and 
$J_A^\mu(x) = {\overline d}(x)\gamma^\mu \gamma^5 u(x)$. Both functions
$\Pi_{V,A}(Q^2\equiv -q^2)$ satisfy the dispersion relation (up to
one subtraction)
\be
\label{dispersion}
\Pi_{V,A}(Q^2)= \int_0^{\infty} \frac{dt}{t+Q^2}\ 
\frac{1}{\pi} {\rm Im}\,\Pi_{V,A}(t)\ .
\ee
Following the discussion in the introduction we will assume that
\be
\label{Imvector}
\frac{1}{\pi} {\rm Im}\,\Pi_V(t)= 2 F_{\rho}^2
\delta(t-M_{\rho}^2) + 2 \sum_{n=0}^{\infty} F_V^2\delta(t-M^2_V(n))\ ,
\ee
\be
\label{Imaxial}
\frac{1}{\pi} {\rm Im}\,\Pi_A(t)= 2 F_{\pi}^2 \delta(t) 
+ 2 \sum_{n=0}^{\infty} F_A^2\delta(t-M^2_A(n))\ .
\ee 
In these expressions, $F_\rho$ is the electromagnetic decay constant
of the $\rho$, and $F_\pi=93$~MeV is the pion decay constant.
$F_V$ and $F_A=F_{a_1}$ are the corresponding constants for the towers of
vector and axial-vector mesons. 
Note that, along with the ansatz of eq.~(\ref{spectrum})
for $M^2_{V,A}(n)$, we take $F_{V,A}$ both to be independent of $n$.

Since \QCD\ has not been solved, it is unclear what values to take for all
these parameters even in the large-$N_c$ limit. We estimate
that $1/N_c$ subleading corrections are roughly of the order of $\Gamma/M\sim
20-30\%$ so that decay constants and masses are uncertain by this amount. 
In addition, we will give results for both $F_{\pi}\simeq 87 $ MeV and 
$F_{\pi}\simeq 93$ MeV as a rough estimate of how much our results change as
a consequence of chiral corrections.

The analysis of this  ansatz is just a straightforward application 
of the Euler-Maclaurin
summation formula. For instance, in the vector case one obtains
\bea
\label{Maclaurin}
\sum_{n=0}^{N_V}\frac{F_V^2}{Q^2+n \Lambda_V^2+m_V^2}&=&
\int_0^{N_V+1} dn \ \frac{F_V^2}{Q^2+n \Lambda_V^2+m_V^2}+ \nonumber \\ 
&+&\frac{1}{2}\ \frac{F_V^2}{Q^2+m_V^2}
+\frac{1}{12}\ \frac{F_V^2 \Lambda_V^2}{(Q^2+m_V^2)^2}+\dots\ ,  
\eea
where $N_V\gg 1$, with the condition that $\Lambda_{\rm cutoff}^2=(N_V+1)
\Lambda_V^2+m_V^2$ is a momentum cutoff which regularizes the theory
and is taken equal to the momentum cutoff in the axial channel in order
to avoid spurious chiral symmetry breaking in the deep ultraviolet.
Performing the integral and expanding in inverse powers of $Q^2$,
one finds
\bea
\label{piV}
\Pi_V(Q^2)&=&\frac{2 F_V^2}{\Lambda_V^2}\log\frac{\Lambda^2_{\rm
    cutoff}}{Q^2} 
+ \frac{1}{Q^2} \left[2 F_{\rho}^2-2 F_V^2
    \left(\frac{m_V^2}{\Lambda_V^2}-\frac{1}{2}\right) \right] \nonumber 
\\ 
&+&\frac{1}{Q^4} \left[-2 F_{\rho}^2 M_{\rho}^2 + F_V^2 \Lambda_V^2 
    \left( \frac{m_V^4}{\Lambda_V^4} - \frac{m_V^2}{\Lambda_V^2}+\frac{1}{6}
\right) \right]\\ \nonumber 
&+& \frac{1}{Q^6} \left[2 F_{\rho}^2 M_{\rho}^4 - \frac{2}{3} F_V^2 
\Lambda_V^4 \left( \frac{m_V^6}{\Lambda_V^6}- \frac{3}{2}
\frac{m_V^4}{\Lambda_V^4} + \frac{1}{2} \frac{m_V^2}{\Lambda_V^2}
\right) \right] + \cO(1/Q^8) \ , 
\eea
and in the axial channel
\bea
\label{piA}
\Pi_A(Q^2)&=&\frac{2 F_A^2}{\Lambda_A^2}\log\frac{\Lambda^2_{\rm
    cutoff}}{Q^2}
+ \frac{1}{Q^2} \left[2 F_{\pi}^2-2 F_A^2
    \left(\frac{m_A^2}{\Lambda_A^2}-\frac{1}{2}\right) \right] \nonumber
\\
&+&\frac{1}{Q^4} \left[F_A^2 \Lambda_A^2
    \left( \frac{m_A^4}{\Lambda_A^4} -
\frac{m_A^2}{\Lambda_A^2}+\frac{1}{6}
\right) \right]\\ \nonumber
&+& \frac{1}{Q^6} \left[- \frac{2}{3} F_A^2
\Lambda_A^4 \left( \frac{m_A^6}{\Lambda_A^6}- \frac{3}{2}
\frac{m_A^4}{\Lambda_A^4} + \frac{1}{2} \frac{m_A^2}{\Lambda_A^2}
\right) \right] + \cO(1/Q^8) \ .
\eea
Of course the expansions of $\Pi_{V,A}(Q^2)$ in eqs.~(\ref{piV},\ref{piA})
have to match their respective OPE expansions \cite{SVZ}. This
leads to the constraints
\be
\label{partonmodel}
\frac{2}{3} \frac{N_c}{16\pi^2}= \frac{F_V^2}{\Lambda_V^2}
= \frac{F_A^2}{\Lambda_A^2} \ ,
\ee
from the coefficient of the parton-model logarithm in front of the unit
operator, 
\be
\label{nodim2}
F_{\rho}^2= F_V^2 \left(\frac{m_V^2}{\Lambda_V^2}-\frac{1}{2}\right) 
\quad {\rm and} \quad
F_{\pi}^2 = F_A^2 \left(\frac{m_A^2}{\Lambda_A^2}-\frac{1}{2}\right)\ ,
\ee
from the absence of a dimension-two operator in the OPE,
\bea
\label{gluoncondensate}
\frac{\alpha_s}{12\pi}\,\langle G_{\mu\nu}G^{\mu\nu}\rangle &=& 
-2 F_{\rho}^2 M_{\rho}^2 + F_V^2 \Lambda_V^2 
    \left( \frac{m_V^4}{\Lambda_V^4} - \frac{m_V^2}{\Lambda_V^2}+\frac{1}{6}
\right)  \nonumber \\
&=& F_A^2 \Lambda_A^2 
    \left( \frac{m_A^4}{\Lambda_A^4} - \frac{m_A^2}{\Lambda_A^2}+\frac{1}{6}
\right) \ ,
\eea
from the contribution of the gluon condensate, and 
\bea
\label{quarkcondensate}
-\frac{28}{9}\pi\alpha_s\langle{\overline \psi}\psi\rangle^2 &=& 
2 F_{\rho}^2 M_{\rho}^4 - \frac{2}{3} F_V^2 
\Lambda_V^4 \left( \frac{m_V^6}{\Lambda_V^6}- \frac{3}{2}
\frac{m_V^4}{\Lambda_V^4} + \frac{1}{2} \frac{m_V^2}{\Lambda_V^2}\right)\ , 
\nonumber \\
 \frac{44}{9}\pi\alpha_s\langle{\overline \psi}\psi\rangle^2 &=& 
- \frac{2}{3} F_A^2 
\Lambda_A^4 \left( \frac{m_A^6}{\Lambda_A^6}- \frac{3}{2}
\frac{m_V^4}{\Lambda_V^4} + \frac{1}{2} \frac{m_A^2}{\Lambda_A^2}\right)\ ,
\eea
from the four-quark condensate.

There are two more constraints which we can impose using
physical observables which depend only on $\Pi_{V,A}(Q^2)$. These are 
the coupling $L_{10}$ of the $\cO(p^4)$ chiral Lagrangian \cite{gasser} 
and the electromagnetic pion-mass difference. They are expressed in
terms of $\Pi_{V,A}(Q^2)$ as
\be
\label{Lten}
L_{10}= -\frac{1}{4} \frac{d}{dQ^2}\left(Q^2\Pi_{LR}(Q^2)\right)_{Q^2=0}\ ,
\ee
and 
\be
\label{pipluspizero}
m_{\pi^+}-m_{\pi^0}= - \frac{3 \alpha}{8 \pi m_{\pi} F_{\pi}^2}
\int_0^{\infty} dQ^2\ Q^2\Pi_{LR}(Q^2)\ ,
\ee
where 
\be
\label{piLR}
\Pi_{LR}(Q^2)\equiv \frac{1}{2}\left(\Pi_V(Q^2)-\Pi_A(Q^2)\right)\ .
\ee

Given $F_{\pi}$, the gluon condensate and the combination
$\pi\alpha_s\langle{\overline \psi}\psi\rangle^2$, we can solve for
the eight parameters $F_{\rho},
M_{\rho}, F_{V,A}, m_{V,A}$ and $\Lambda_{V,A}$ from
eqs.~(\ref{partonmodel}--\ref{quarkcondensate}), and from the solution
determine $L_{10}$ and
$m_{\pi^+}-m_{\pi^0}$. Thus, the system is overconstrained. 

\section{Phenomenological results}

In principle, one could solve the system of equations 
eqs.~(\ref{partonmodel}-\ref{quarkcondensate}) as indicated 
at the end of the previous section.  However, the condensates are
relatively poorly known, and we prefer to follow a different approach.
We will use as inputs the values of $F_\pi=93$~MeV (we will also
consider the value $F_\pi=87$~MeV, which is the value in the chiral
limit \cite{gasser}), $M_\rho=770$~MeV\footnote{Since one should actually use
  large-$N_c$ values, one could try to vary the $\rho$ mass as
  well. However, since our choice $M_\rho\simeq 770$~MeV already does a good
  job, we decided to keep the analysis as simple as possible and {\it not} to
  fiddle with $M_\rho$, while we wait for more data to come in (for instance on
  the spectrum).}, and $F_\rho$ ranging from
about $120$ to $150$~MeV, the latter being the experimental value.
(For larger values the combination 
$\pi\alpha_s\langle{\overline \psi}\psi\rangle^2$ becomes negative,
in contradiction with the phenomenologically most favored value. See below.)
Choosing a range like this is reasonable, because we are in the large-$N_c$
limit, and expect variations of order $20-30$\%.  It allows us to
eliminate the condensates from eqs.~(\ref{gluoncondensate},\ref{quark%
condensate}), and solve the remaining two equations 
along with eqs.~(\ref{parton%
model},\ref{nodim2}) for $F_{V,A}$, $m_{V,A}$, and $\Lambda_{V,A}$,
and then calculate the two condensates in terms of these.

The solutions for $F_{V,A}$ are shown in figure 1, those
for $m_{V,A}$ in figure 2, and the condensates are shown in
figures 3 and 4, as a function of $F_\rho$.  Below $F_\rho\approx 123$~MeV 
the equations do not have a real solution.  In the axial channel,
$m_A=M_{a_1}=1230$~MeV experimentally, while for $F_A=F_{a_1}$ 
a value of $135$~MeV is reported in ref.~\cite{ecker}.  $F_V$ is not
known experimentally.  Phenomenological estimates for the condensates are
$\alpha_s\langle G^2\rangle=0.048\pm 0.030$~GeV$^4$ \cite{yndurain},
and $\pi\alpha_s\langle{\overline\psi}\psi\rangle^2=9\pm 2
\times 10^{-4}$~GeV$^6$ \cite{davier}.  Looking at figures
1 to 4, we see that very good agreement is obtained for values of
$F_\rho$ of about $130-135$~MeV.  For $F_\rho=130$~MeV we summarize
both our solution along with the experimental values in table 1,
in which we also give our results for $F_\pi=87$ instead of $93$~MeV.

\begin{figure}[bht]
\begin{center}
\epsfig{file=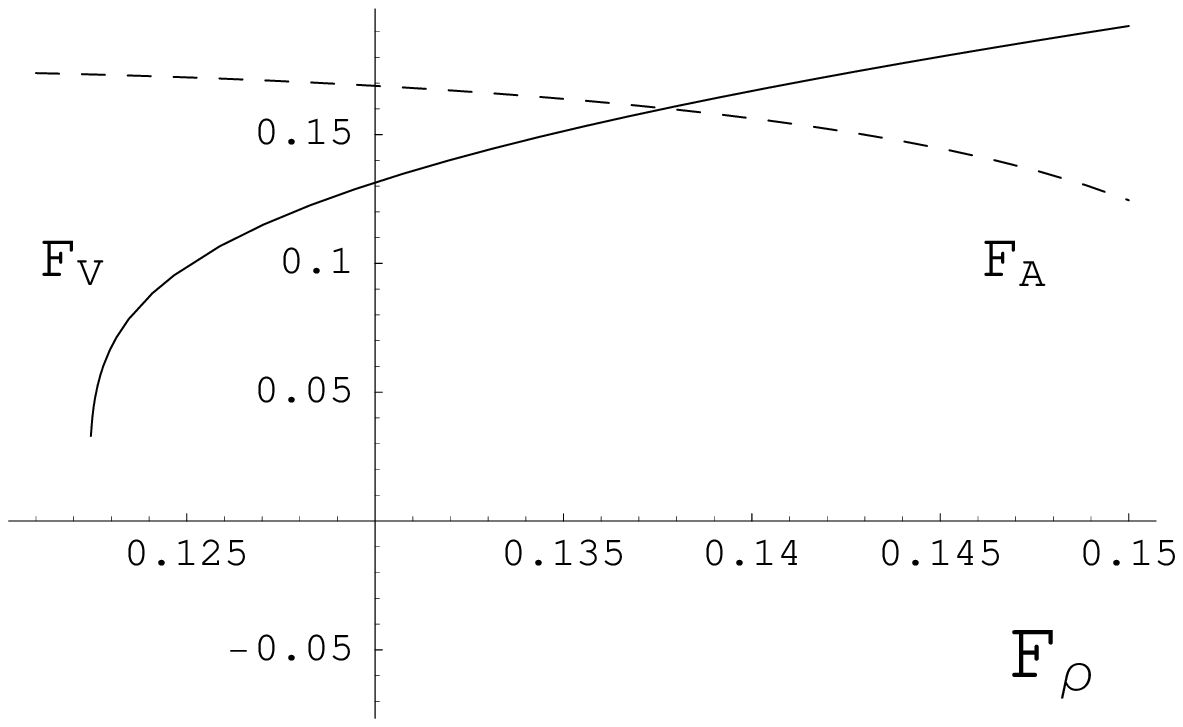,width=10cm,height=8cm} 
\end{center}
\vskip -1.5cm
\caption{}{Decay constants $F_V$, $F_A$ in eqs.~(\ref{Imvector},\ref{Imaxial}) 
as a function of $F_{\rho}$ in eq.~(\ref{Imvector}). All in GeV.}
\end{figure}

\begin{figure}[bht]
\begin{center}
\epsfig{file=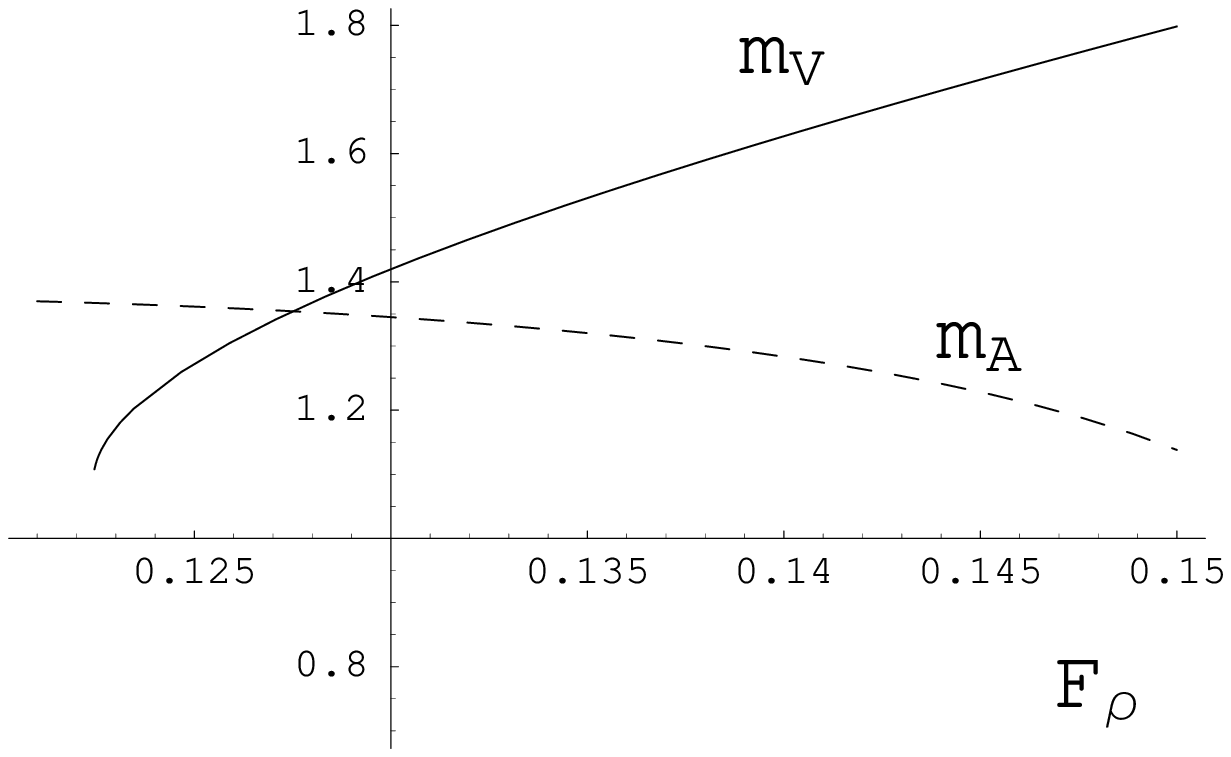,width=10cm,height=8cm} 
\end{center}
\vskip -1.5cm
\caption{}{Resonance masses in eq.~(\ref{spectrum}) as a function of 
$F_{\rho}$ in eq.~(\ref{Imvector}). All in GeV. }
\end{figure}

\begin{figure}[bht]
\begin{center}
\epsfig{file=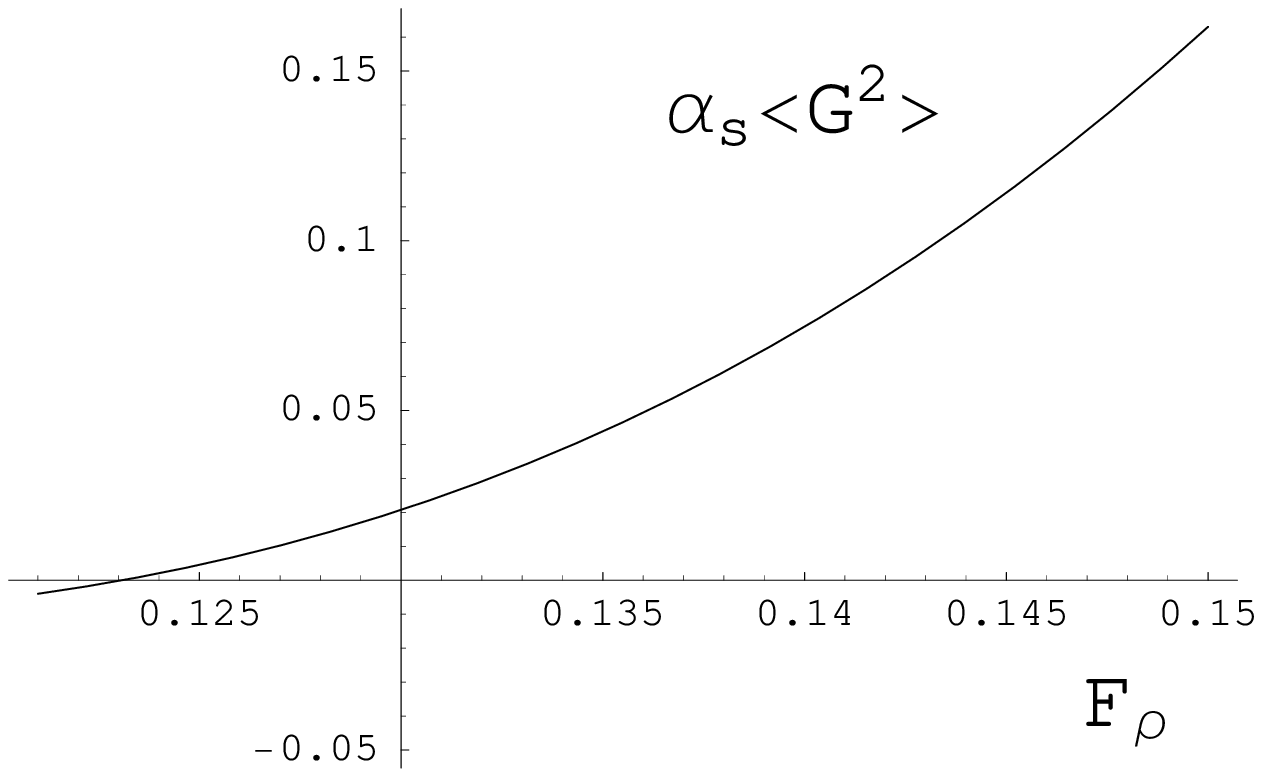,width=10cm,height=8cm} 
\end{center}
\vskip -1.5cm
\caption{}{The gluon condensate $ \alpha_s\langle G^2\rangle$ (GeV$^4$) as a
  function of $F_{\rho}$ (GeV) in eq.~(\ref{Imvector}). }
\end{figure}

\begin{figure}[bht]
\begin{center}
\epsfig{file=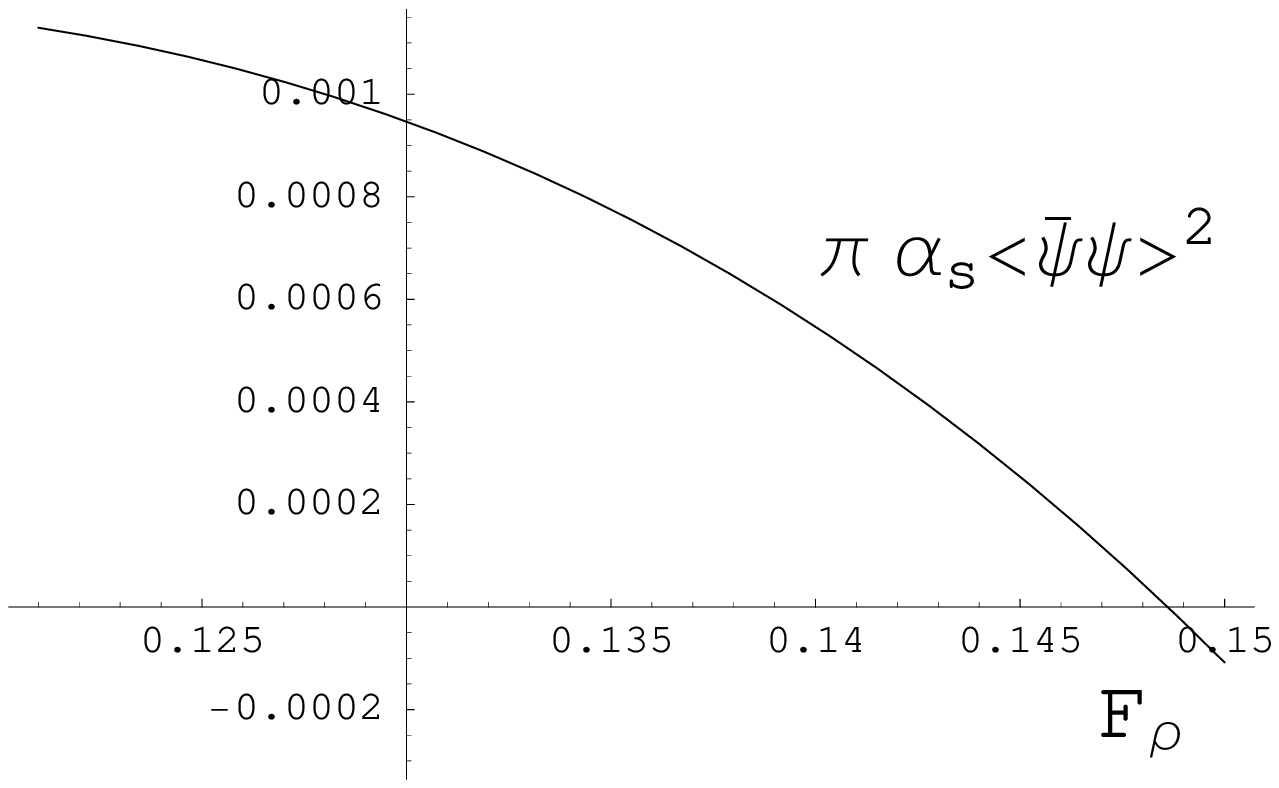,width=10cm,height=8cm} 
\end{center}
\vskip -1.5cm
\caption{}{The four quark condensate 
$\pi \alpha_s\langle{\overline \psi}\psi\rangle^2$(GeV$^6$) as a
  function of $F_{\rho}$ (GeV) in eq.~(\ref{Imvector}). }
\end{figure}

\begin{table}
\caption[Results]{Results obtained for two different values of $F_{\pi}$. 
$M_{\rho}=770$~MeV is input, and we choose $F_\rho=130$~MeV. }
\vskip 0.5cm
\begin{tabular}{cccc}
\hline
\hline 
        & $F_{\pi}=$ 87 MeV & $F_{\pi}=$93 MeV & Experiment\\ \hline
$F_{\rho}$ (MeV) &   130     &  130   &$153\pm 4$ \cite{pdg}\\ \hline
$M_{\rho}$ (GeV)&   0.77    & 0.77   & $0.7700\pm 0.0008$ \cite{pdg} \\ \hline
$ F_V$ (MeV)    &   140     &  130   &  no available data     \\ \hline
$ m_V$ (GeV)    &   1.44    & 1.42  & $1.465\pm 0.025$ \cite{pdg}  \\ \hline
$ \Lambda_V$ (GeV) &  1.21    & 1.17       & see Table 2 \\ \hline
$F_{a_1}= F_A$ (MeV)    &   160   & 170     &  $135\pm 30$ \cite{ecker} \\ \hline
$M_{a_1}= 
m_A$ (GeV)    &   1.27 & 1.35   &  $1.230\pm 0.040$ \cite{pdg}\\ \hline
$ \Lambda_A$ (GeV)  & 1.42   & 1.50   &  see Table 2\\ \hline
$ \alpha_s\langle G^2\rangle$ (GeV$^4$) & 0.01 & 0.02 & 
$0.048\pm 0.030$ \cite{yndurain}  
\\ \hline
$\pi \alpha_s\langle{\overline \psi}\psi\rangle^2$($10^{-4}$GeV$^6$)& 
7 & 9 & $(9\pm 2)$
\cite{davier} \\ \hline
$ m_{\pi^+}-m_{\pi^0}$ (MeV) & 4.2 & 4.7 & $4.5936\pm0.0005 $ \cite{pdg}  
 \\ \hline
$ L_{10}(\mu=M_{\rho})$ ($10^{-3}$) & $-5.2$  & $-5.6$  & $-5.13\pm
0.19$ \cite{davier}   
\\ \hline \hline

\end{tabular}
\label{table1}
\end{table}

We may now use the values obtained for $m_{V,A}$ and $\Lambda_{V,A}$
in order to calculate the vector and axial-vector meson spectrum,
from eq.~(\ref{spectrum}).  
The results are shown in table 2.  While experimental 
confirmation for the mass of the 
$a_1'$ (and higher resonances) is needed, 
table 2 shows a good agreement between the model and the real world. 

\begin{table}
\caption[Results]{Vector and axial-vector mass spectra in GeV with the
corresponding experimental values, where available.}
\vskip 0.5cm
\begin{tabular}{cccccccc}
\hline
\hline 
                 & $\rho'$ & $\rho''$& $\rho'''$ & $\rho''''$
& $a_1  $  & $a'_1$  & $a''_1$\\ \hline
$F_{\pi}=87$ MeV & 1.4      &  1.9     &   2.2      &  2.5  
& 1.3     &  1.9     &  2.4        \\ \hline
$F_{\pi}=93$ MeV & 1.4      &  1.8     &   2.2      &  2.5  
& 1.3     &  2.0     &  2.5        \\ \hline
Exp. \cite{pdg,baker,dorofeev} &  1.47   &  1.70    &  2.15   & no data  
& 1.23  & 1.64 \cite{baker}, 1.8 \cite{dorofeev}  &  no data  \\ \hline
\hline
\end{tabular}
\label{table2}
\end{table}

Using these results, eq. (\ref{spectrum}) can now be rewritten as
\be
\label{daughter}
M^2_V(n)=M^2_{\rho'}\left( 1 + a\ n\right)\quad ,\quad  a\simeq 0.7 \quad .
\ee
This is actually very close to the original proposal of
ref. \cite{bramon,sakurai} where the $\rho$ is not a separate resonance from
the equally spaced tower but just another member. In order to  make the 
comparison more clear, 
rewrite eq. (\ref{GVMD}) in terms of the $\rho'$ mass instead of the $\rho$.
Equation (\ref{GVMD}) (with $\alpha=2$) then reads 
$M_V^2(n)= M^2_{\rho'} ( 1 + 2n/3)$ and $a=2/3$, to be compared with
eq. (\ref{daughter}).  
However, although the spectrum is quite 
similar, other quantities are not. As a matter of fact, 
note that the result $\alpha=2$ in eq. (\ref{GVMD})
(i.e. $m_V^2/\Lambda_V^2=1/2$), 
can {\it also} be derived as a consequence of the lack of
dimension-two operator in the OPE: the solution $\alpha=2$ follows from 
eq. (\ref{nodim2}) (after
setting  $F_{\rho}=0$, since the $\rho$ meson belongs to the equally spaced 
tower of refs. \cite{bramon,sakurai}).   However, 
eq. (\ref{gluoncondensate}) then yields a negative gluon condensate
(the gluon condensate is very sensitive to variations in meson masses
and decay constants).

{}From eq.~(\ref{Lten}), we find for the ES  ansatz
\be
\label{Ltenexp}
L_{10}=-\frac{1}{4}\left(\frac{F^2_\rho}{M^2_\rho}
+\left(\frac{m^2_A}{\Lambda^2_A}-\frac{m^2_V}{\Lambda^2_V}\right)
\sum_{n=0}^\infty\frac{1}{(n+\frac{m^2_V}{\Lambda^2_V})
(n+\frac{m^2_A}{\Lambda^2_A})}+
\frac{2}{3}\frac{N_c}{16\pi^2}\log\frac{F^2_A}{F^2_V}\right)\ .
\ee
For $F_\rho=130$~MeV the values we obtain are recorded in table 1.
The value of $L_{10}$ varies by less than 15\% over the range of
$F_\rho$ we considered.

For $m_{\pi^+}-m_{\pi^0}$ the expression following from 
eq.~(\ref{pipluspizero})
is rather lengthy, and we relegate it to an appendix.  Again, the
values for $F_\rho=130$~MeV are presented in table 1.  
In the calculation of the pion-mass difference we take $M_\pi=135$~MeV.
At $F_\rho=150$~MeV, we find that $m_{\pi^+}-m_{\pi^0}=3.0$~resp.~$3.2$~MeV
for $F_\pi=87$~resp.~$93$~MeV.  It is clear that the ES  ansatz favors
a lower value of $F_\rho$ of around $130$~MeV.  Since we find no
solution for $F_\rho\le 120$~MeV, the solution is quite constrained.

In principle, our model may be used to predict hitherto unknown values of 
quantities of phenomenological interest.  Since we have restricted our
attention to vector and axial-vector two-point functions, there are
not many such quantities.  One such quantity is a $K\to\pi$ matrix element
of the electro-magnetic penguin operator $Q_7$,\footnote{The 
physical $K\to(\pi\pi)_{I=2}$ matrix element is related to this simpler matrix
element by ChPT.}
which can be expressed solely in terms of $\Pi_{LR}$ \cite{knechtetal}:
\be
\label{Qseven}
Q^{3/2}_7(\mu)\equiv
\langle\pi^+|Q^{3/2}_7|K^+\rangle(\mu)=\frac{3}{8\pi^2 f^2_\pi}
\int_0^{\mu^2}dQ^2\,Q^4\Pi_{LR}(Q^2)\ .
\ee
Even though this quantity by itself is of limited phenomenological
interest, it is interesting to compare the value we obtain  with
that obtained from Lowest-Meson Dominance \cite{previouswork}
 (LMD; {\it cf.} section 4
below).  The difference in numerical values gives an indication of
the robustness of this sum-rule approach with respect to variations
in the  ansatz taken for the spectral functions.  In addition,
lattice \cite{lattice} and other phenomenological estimates 
\cite{donoghue,narison} exist \cite{announcement}.
Note that this quantity is scale dependent, because the integral
is logarithmically divergent.  This is because terms up to $1/Q^4$
cancel in $\Pi_{LR}$ ({\it cf.} eq.~(\ref{piLR})), but not terms of
order $1/Q^6$.  
We will return to this quantity in the next section.

\section{Comparison with the Lowest-Meson Dominance approximation}

It is interesting to compare the ES  ansatz which we have been
considering thus far with the Lowest-Meson Dominance
(LMD)  ansatz of ref. \cite{previouswork}.  In the ES case, the
(leading) perturbative values of $\Pi_{V,A}(Q^2)$ for large $Q^2$
come from the assumed Regge-like spectrum.  One may also directly
incorporate pQCD into an  ansatz by approximating $\Pi_{V,A}(Q^2)$
with the expression one finds in pQCD which, 
to leading order in $\alpha_s$, reads 
\be
\label{continuum}
\frac{1}{\pi} {\rm Im}\,\Pi_V(t)|_{pQCD}= \frac{2}{3}\frac{N_c}{16 \pi^2}
\,\theta
\left(t-s_0\right) \ .
\ee
Here a parameter $s_0$ is introduced which determines the onset of
the perturbative continuum, above which we expect pQCD to give accurate
results.  For $t\ge s_0$ this expression replaces the sums in 
eqs.~(\ref{Imvector},\ref{Imaxial}).  All resonances
for $t<s_0$ are kept;  for an illustration, see figure 5.
In the simplest version of this approach, one keeps one resonance
each in the vector and axial-vector channels.  Since, by definition,
$s_0$ in eq.~(\ref{continuum}) marks the onset of the
perturbative continuum, it is taken to be the same for both vector
and axial-vector channels, because chiral symmetry is not broken in
perturbation theory. This  ansatz 
has recently been shown to yield a successful
description of Aleph data for the moments of the $\Pi_{LR}$
function \cite{phily}, it successfully predicts pseudoscalar
decays \cite{pseudo} and, also recently, it has allowed an analytic calculation
of the $K^0-\bar K^0$ mixing parameter, $B_K$ \cite{BK}.

\begin{figure}[bht]
\begin{center}
\epsfig{file=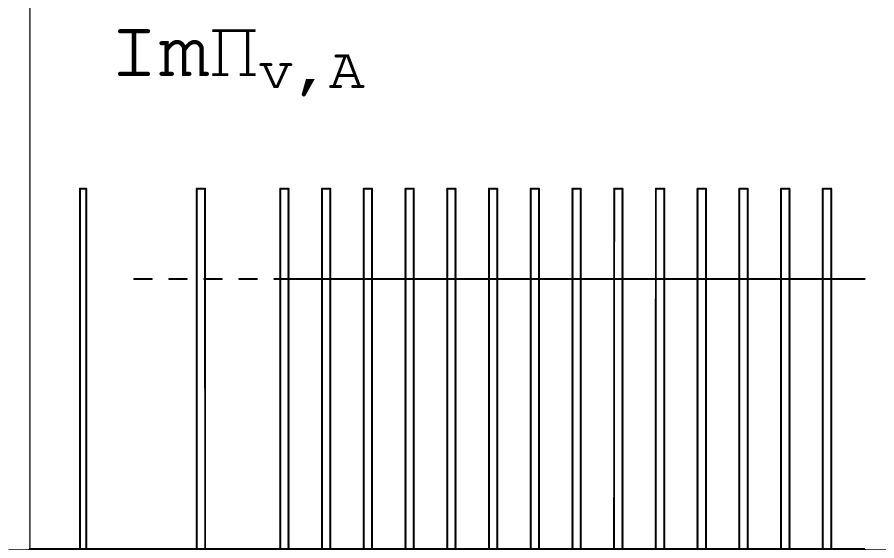,width=10cm,height=8cm} 
\end{center}
\vskip 0cm
\caption{}{Schematic view of the spectrum (set of vertical lines)  
together with the parton model, eq.~(\ref{continuum})
(flat horizontal line). }
\end{figure}

Qualitatively one can understand the transition between ES and LMD in the
following way. 
Low-energy observables are dominated by the lowest individual resonances to
the extent that the effect of the continuum is suppressed by the
fact that $s_0$ is larger than the mass (squared) of these individual
resonances. For the LMD case, the continuum (\ref{continuum}) produces a
logarithm just like the integral in eq. (\ref{Maclaurin}) does, with 
$s_0=24\pi^2F_\rho^2/N_c=m_V^2$ (see refs. \cite{previouswork,seveneleven} 
for details on the LMD ansatz).
Taking also into account the leading correction term in
eq.~(\ref{Maclaurin}) to order $1/Q^2$, we find $s_0=m_V^2\left(1-\frac{1}{2}
\Lambda_V^2/m_V^2\right)$.  Taking $\Lambda_V^2/m_V^2$ smaller
(keeping $F_V^2/\Lambda^2_V$ fixed, {\it cf.} eq.~(\ref{partonmodel}))
clearly approaches the perturbative approximation of
eq.~(\ref{continuum}), with $s_0\to m_V^2$, and with the correction terms
on the right-hand side of eq.~(\ref{Maclaurin}) becoming smaller
(for any value of $Q^2$).  In this sense, the LMD approach can be
seen as following from the ES approach in the limit 
$\Lambda_V^2/m_V^2\to 0$, at least in the vector channel.  (In the
axial channel, a similar argument would apply if we had taken one
separate resonance apart from the tower in the ES approach.
In our case, $m_A$ is relatively small, and $\frac{1}{2}\Lambda^2_A/m^2_A$
is substantially bigger than $\frac{1}{2}\Lambda^2_V/m^2_V$.)

Of course, it is easy to put 
$\alpha_s$ corrections into eq.~(\ref{continuum}) in the LMD approach,
while this is less straightforward in the ES approach (it requires
a more complicated $n$ dependence of $M^2_{V,A}$ and $F^2_{V,A}$).
One can also improve the extent to which LMD approximates ES by 
keeping more individual resonances explicit in the LMD  ansatz.
This will move up $s_0$ or equivalently $m_V^2$, hence reducing the
size of $\Lambda_V^2/m_V^2$.

Using the same information from the OPE, {\it and} the relation
$F_\rho^2=2F_\pi^2$ \cite{swisscatalan}, one finds that in the LMD
approach $s_0$ is determined to be $s_0=(4\pi F_\pi)^2$ from the
vector channel, and then, from the axial-vector channel, that
$M_{a_1}$ is close to, but smaller than $\sqrt{s_0}$ \cite{seveneleven}.%
\footnote{In comparing
with ref.~\cite{seveneleven}, note that $F_\rho=f_\rho M_\rho$,
{\it etc.}}  
This implies that within LMD the $a_1$ resonance
has to be kept, and cannot be folded into the perturbative
continuum, eq.~(\ref{continuum}).  Higher resonances are well
above $\sqrt{s_0}$ (whether we take their experimental values, or
those predicted from ES, {\it cf.} table 2).  We will therefore
compare the phenomenology of the ES  ansatz, summarized
in tables 1 and 2, with that of the LMD ansatz with one
explicit resonance in each channel (in addition to the pion).  
The LMD results are presented
in table 3.  All quantities displayed in table 3 are 
predicted in terms of $F_\pi$ with the LMD  ansatz
\cite{previouswork,seveneleven}, using the relation $F_\rho^2=2F_\pi^2$%
, but (obviously) no predictions
are made for excited states in both channels in this case.  In the comparison
with the results in table 1, one should keep in mind that the
relation $F_\rho^2=2F_\pi^2$ was not used with the ES  ansatz.
However, for $F_\pi=87$, resp. $93$~MeV, $F_\rho=130$~MeV satisfies this
relation at the $10$, resp. $2$\% level.

Comparing the results of ES and LMD, it is clear that both
approaches yield consistent results, especially if one takes
into account that all results are in the large-$N_c$ and chiral
limits, so that deviations of less than $30$\% are not significant.
Possible exceptions to this are 
$F_{a_1}$ and  $\pi \alpha_s\langle{\overline \psi}\psi\rangle^2$, 
for which the LMD values are   
smaller, resp. larger, than the ES value.  Unfortunately, the experimental
values, while closer to the ES values, are rather poorly known.
Possibly with these two exceptions, the differences between tables
1 and 3 give a good indication of the spread one can expect
when one combines QCD techniques such as large $N_c$, OPE and
pQCD with an  ansatz for the spectrum
in order to obtain explicit solutions.

Finally, we compare the ES prediction for $Q_7^{3/2}(\mu)$ with
the LMD prediction for the same number.  First, we show the function
$Q^4\Pi_{LR}(Q^2)$ in figure 6, for both the ES and LMD
ans\"atze.  We evaluated the ES value of
$Q_7^{3/2}(\mu)$ at $\mu=2$~GeV from the area under the curve using $\mu$ as a
sharp cutoff 
(choosing $F_\rho=130$~MeV and $M_\rho=770$~MeV),
and quote the LMD value from ref.~\cite{knechtetal},
finding, 
\be
\label{qvalues}
Q_7^{3/2}(\mu=2\,{\rm GeV})=
\cases{-0.021\ ({\rm GeV})^4\,,&ES,\ \ \ \ $F_\pi=93$~MeV,\cr
-0.019\ ({\rm GeV})^4\,,&ES,\ \ \ \ $F_\pi=87$~MeV,\cr
-0.024\ ({\rm GeV})^4\,,&LMD.}
\ee
We took again $F_\rho^2=2F_\pi^2$ in the LMD case, in which case
the LMD value is independent of $F_\pi$ (for a fixed $M_\rho$).  
We see that, as in the case of
most other quantities, the predicted
value we obtain for $Q_7^{3/2}(\mu=2\,{\rm GeV})$ is quite
robust to the variation of the  ansatz for the vector
and axial-vector spectral functions we have been considering.
The values we obtain are fully consistent with each other
within the context of the large-$N_c$ and chiral limits.

\begin{figure}[bht]
\begin{center}
\epsfig{file=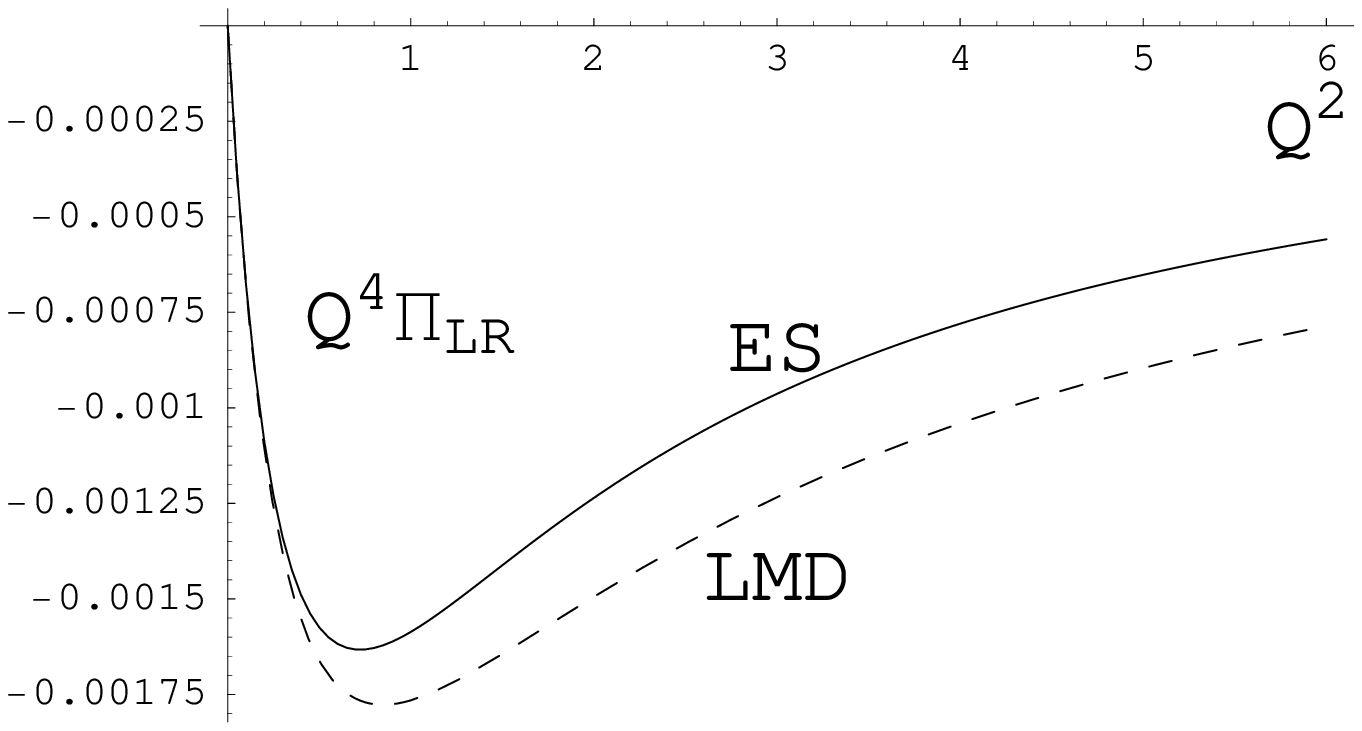,width=10cm,height=8cm} 
\end{center}
\vskip -1.5cm
\caption{}{The function $Q^4 \Pi_{LR}(Q^2)$ (GeV$^4$) for our equal-spacing
  model (ES) (solid line) and its lowest meson dominance version (LMD) (dashed
  line) as a function of the euclidean momentum $Q^2$ (GeV$^2$).}
\end{figure}

\begin{table}
\caption[Results]{LMD results obtained for two different values of
$F_{\pi}$.  }
\vskip 0.5cm
\begin{tabular}{ccc}
\hline
\hline
& $F_{\pi}=$ 87 MeV & $F_{\pi}=$93 MeV \\ \hline
$F_{\rho}$ (MeV) &   122     &  131   \\ \hline
$M_{\rho}$ (GeV)&   0.77    & 0.82   
\\ \hline
$F_{a_1}$ (MeV)    &   87   & 93      \\
\hline
$M_{a_1}$ (GeV)    &   1.08 & 1.16   \\
\hline
$ \alpha_s\langle G^2\rangle$ (GeV$^4$) & 0.01 & 0.02 
\\ \hline
$\pi \alpha_s\langle{\overline \psi}\psi\rangle^2$($10^{-4}$GeV$^6$)&
13 & 19 \\ \hline
$ m_{\pi^+}-m_{\pi^0}$ (MeV) & 5.2 & 6.0
\\ \hline
$ L_{10}(\mu=M_{\rho})$ ($10^{-3}$) & $-4.8$  & $-4.8$ 
\\ \hline \hline

\end{tabular}
\label{table3}
\end{table}


\section{Conclusion}

In this paper we have considered the vector and
axial-vector current two-point functions in QCD.  The combination 
of taking the large-$N_c$ and chiral limits, the use of 
asymptotic freedom and the OPE
constrain the possible form of these two-point functions.
However, this is 
not enough to obtain concrete predictions: more knowledge,
or lacking that, an  ansatz for the spectral functions in
these two channels is still needed.  

A particular  ansatz which has been considered in the
literature is the LMD  ansatz, in which the high-energy
part of the spectral functions is calculated in pQCD, while
one (or a few) explicit resonances are kept at low to medium energy.
This approach has been quite successful in reproducing known
phenomenology, and should therefore be useful in estimating
as yet unknown hadronic quantities, such as weak matrix elements.
However, although LMD yields a rather intuitive picture it is not a simple
matter to give it a fully systematic treatment. Such systematic treatment
should, in particular, produce a theory of the so-called quark-hadron duality
violations \cite{shifman}, a problem which is clearly beyond the scope of the
present work. Consequently,  
determinations of the systematic errors associated with this approach have to
rely at some point on phenomenological experience.

One way to address this issue is to vary the nature of the
ansatz within the confines of our general picture of
the strong interactions, and this is what we have done in this
paper.  Instead of LMD, we have based our  ansatz
on Regge theory, except for the lowest states in each channel
(the $\rho$ meson and the pion).  For the precise form of
the ES  ansatz, see eqs.~(\ref{Imvector},\ref{Imaxial}).
The hadron phenomenology calculated with this assumption
is again very good, and agrees well with what is found with
LMD ({\it cf.} tables 1 and  3).  
Possible exceptions perhaps are  $F_{a_1}$ and 
$\pi \alpha_s\langle{\overline \psi}\psi\rangle^2$;
experimentally these quantities are not that well-known.  
In general, because of the use of large-$N_c$ and
chiral limits, agreement at the 20-30\% level, between ES
and LMD, or with experiment, should be considered satisfactory.

The gluon and four-quark condensates are rather sensitive to
the values of other quantities (masses and decay constants).
Within our solution ({\it cf.} table 1),
we find that the $\rho$ comes out to be very close to
lying on the daughter trajectory predicted by Regge theory 
in that $M^2_{\rho'}\approx M^2_\rho+\Lambda^2_V$ and
$F_\rho\approx F_V$.  However, requiring the $\rho$ to be {\it precisely} 
on this trajectory leads to a negative gluon condensate.
Since large $N_c$ (supplemented with Regge Theory) 
and pQCD suggest that asymptotically
the tower of resonances has to be equally spaced, it makes
intuitive sense to model the resonance spectrum with an
equally spaced tower above $1$~GeV, keeping those below
$1$~GeV separate.  The ES  ansatz goes beyond LMD in
that it yields a value for the spacing between the resonances
(the parameters $\Lambda_{V,A}$), and leads to a reasonable
resonance pattern in the vector and axial-vector channels ({\it cf.} table 2).

We have then used the ES ansatz in order to predict
a weak matrix element, $Q^{3/2}_7(\mu)$, which is of interest
in the context of non-leptonic kaon decays.  We find a value
rather close to that predicted from LMD, thus lending support
to the credibility of this prediction. 

Finally, we discussed the extent to which the two 
ans\"atze are equivalent.  We qualitatively argued that in the limit
in which we take the spacing between the resonances in the ES
tower to zero one recovers the LMD approach.  However,
with LMD, we kept both the $\rho$ and the $a_1$ as separate
resonances (because one finds that the perturbative
threshold $s_0$ is larger than $M^2_{a_1}$), while we did
take the $a_1$ to be the first resonance of the ES tower. 
  
After all these years we still do not have a real understanding of the
large-$N_c$ expansion in QCD. Clearly the ES model is not meant to be a
substitute for large-$N_c$ QCD at the theoretical level but, given its
phenomenological success, it may help in our understanding of this limit of
QCD. 

\vskip 1 cm

\leftline{\bf Acknowledgments}

\vspace*{3mm}

\noi
MG and SP thank E. de Rafael for useful discussions and M. Knecht for
reading the manuscript. SP thanks A. Bramon for interesting 
conversations on Regge Theory and 
F. Jegerlehner for drawing his attention to ref. \cite{geshkenbein}
in the course of this work. MG and SP thank the IFAE at the
Universitat Aut\`onoma de Barcelona and the Department of Physics
at Washington University for mutual hospitality.  MG was
supported in part by the US Dept. of Energy and by a fellowship
of the Spanish Government SAB1998-0171. SP was supported by
CICYT-AEN99-0766 and by TMR, EC-Contract No. ERBFMRX-CT980169 (EuroDaphne).

\section*{Appendix}

The expression for the electromagnetic pion mass difference,
following from the ES  ansatz follows from eq.~(\ref{pipluspizero}),
is
\bea
\label{mpidiff}
m_{\pi^+}-m_{\pi^0}&=&
\frac{3\alpha}{8\pi m_\pi f^2_\pi}
\Biggl\{-\frac{1}{2}F_\rho^2 M^2_\rho+\frac{1}{2}F_A^2 m^2_A \nn \\
&&-\frac{N_c}{24\pi^2}
\Bigg[\frac{11}{12}m^4_V\log{\frac{M^2_\rho}{m^2_V}}
-\frac{7}{12}(m^2_V+\Lambda^2_V)^2
\log{\frac{M^2_\rho}{m^2_V+\Lambda^2_V}}
\nn\\
&&\qquad \qquad +\frac{1}{6}(m^2_V+2\Lambda^2_V)^2
\log{\frac{M^2_\rho}{m^2_V+2\Lambda^2_V}}\Bigg] \nn\\
&&+\frac{N_c}{24\pi^2}
\Bigg[\frac{11}{12}(m^2_A+\Lambda^2_A)^2\log{\frac{m^2_A}
{m^2_A+\Lambda^2_A}}\nn\\
&&-\frac{7}{12}(m^2_A+2\Lambda^2_A)^2
\log{\frac{m^2_A}{m^2_A+2\Lambda^2_A}}
+\frac{1}{6}(m^2_A+3\Lambda^2_A)^2
\log{\frac{m^2_A}{m^2_A+3\Lambda^2_A}}\Bigg]\nn \\
&&+\frac{1}{2}\left[2F^2_\rho M^2_\rho
-\frac{24\pi^2}{N_c}F^4_\rho
+\frac{1}{12}F^2_V\Lambda^2_V\right]\log\frac{m^2_A}{M^2_\rho} \nn\\
&&
-3F^2_V\Lambda^2_V G\left(\frac{m^2_V}{\Lambda^2_V}\right)
+3F^2_A\Lambda^2_A G\left(\frac{m^2_A}{\Lambda^2_A}\right)\Biggr\}\ ,
\eea
where
\bea
\label{G}
G(x)&=&\sum_{n=0}^\infty\Biggl(\frac{1}{(n+x)^2}
+4(n+1+x)^3\log\frac{n+x}{n+1+x}
-3(n+2+x)^3\log\frac{n+x}{n+2+x} \nn \\
&& +\frac{4}{3}(n+3+x)^3\log\frac{n+x}{n+3+x}
-\frac{1}{4}(n+4+x)^3\log\frac{n+x}{n+4+x}\Biggr)\,.
\eea
This expression results from calculating the integral in 
eq.~(\ref{pipluspizero}), which is finite, because
$\Pi_{LR}(Q^2)\sim 4\pi\alpha_s\langle\overline\psi\psi\rangle^2/
Q^6$ for large $Q^2$.  By partial integration 
\bea
\label{detail}
\int_0^\infty dQ^2\,Q^2\Pi_{LR}(Q^2)&=&
-\frac{1}{2}\int_0^\infty dQ^2\,Q^4\frac{d\ }{dQ^2}\Pi_{LR}(Q^2)\ .
\eea
The sums over $n$ inside $\frac{d\ }{dQ^2}\Pi_{LR}(Q^2)$
now converge.  Next, one would like to make use of the fact
that terms up to order $1/Q^8$ in $\frac{d\ }{dQ^2}\Pi_{LR}(Q^2)$
vanish, as follows from
eqs.~(\ref{partonmodel}--\ref{gluoncondensate}).
This can be done by writing
\bea
\label{sums}
\sum_{n=0}^\infty\frac{F_V^2}{(Q^2+m_V^2+n\Lambda^2_V)^2}
&=& \\
&&\hspace{-2truecm}\sum_{n=0}^\infty\Biggl(
\frac{F_V^2}{(Q^2+m_V^2+n\Lambda^2_V)(Q^2+m_V^2+(n+1)\Lambda^2_V)}\nn\\
&&
+\frac{F_V^2}{(Q^2+m_V^2+n\Lambda^2_V)^2
(Q^2+m_V^2+(n+1)\Lambda^2_V)}\Biggr) \nn \\
=\frac{F_V^2}{\Lambda_V^2(Q^2+m_V^2)} &+&\sum_{n=0}^\infty
\frac{F_V^2}{(Q^2+m_V^2+n\Lambda^2_V)^2
(Q^2+m_V^2+(n+1)\Lambda^2_V)}\ ,  \nn
\eea
using
\bea
\label{formula}
\sum_{n=0}^N\frac{1}{(p+nq)(p+(n+1)q)\dots(p+(n+\ell)q)}&=& \\
\frac{1}{\ell q}\Biggl(\frac{1}{p(p+q)\dots(p+(\ell-1)q)}
&-&\frac{1}{(p+(N+1)q)\dots(p+(N+\ell)q)}\Biggr) \nn
\eea
with $N=\infty$ and $\ell=1$.  This process can be repeated
by employing this equation for higher values of $\ell$ until
the ``left-over" sum on the right-hand side of eq.~(\ref{sums})
is of order $1/Q^8$, so that the integral over $Q^2$ in
eq.~(\ref{detail}) can be done term by term.  The remaining
terms combine with similar terms in the axial channel, and
the $Q^2$ integral over these combined terms is finite as well,
by virtue of eqs.~(\ref{partonmodel}--\ref{gluoncondensate}).
One ends up with the expression given in eqs.~(\ref{mpidiff},\ref{G}).

\end{document}